# Effect of *In-vitro* Heat Stress Challenge on the function of Blood Mononuclear Cells from Dairy Cattle ranked as High, Average and Low Immune Responders

Shannon L Cartwright[1], Marnie McKechnie[1], Julie Schmied[1], Alexandra M Livernois[1,2] and Bonnie A Mallard[1,2]

1. Department of Pathobiology, Ontario Veterinary College, University of Guelph, 50 Stone Rd, Guelph, ON., N1G 2W1, Canada.
2. Centre of Genetic Improvement of Livestock, Animal Biosciences, University of Guelph, 50 Stone Rd, Guelph, Ont., NIG 2W1, Canada

Corresponding author: S.L. Cartwright- cartwris@uoguelph.ca

**Abstract**

**Background:** The warming climate is causing livestock to experience heat stress at an increasing frequency. Holstein cows are particularly susceptible to heat stress because of their high metabolic rate. Heat stress negatively affects immune function, particularly with respect to the cell-mediated immune response, which leads to increased susceptibility to disease. Cattle identified as having enhanced immune response have lower incidence of disease. Therefore, the objective of this study was to evaluate the impact of *in vitro* heat challenge on blood mononuclear cells from dairy cattle, that had previously been ranked for immune response, in terms of heat shock protein 70 concentration, nitric oxide production, and cell proliferation.

**Methods:** Bovine blood mononuclear cells, from Holstein dairy cattle previously ranked for immune response based on their estimated breeding values, were subjected to three heat treatments: thermoneutral, heat stress 1 (incubate cells for one heat treatment at 42℃) and heat stress 2 (incubate cells for two heat treatments at 42℃). Cells of each treatment were evaluated for heat shock protein 70, cell proliferation and nitric oxide production.

**Results:** Blood mononuclear cells from dairy cattle classified as high immune responders, based on their estimated breeding values for antibody and cell-mediated responses, produced a significantly greater concentration of heat shock protein 70 under most heat stress treatments compared to average and low responders, and greater cell-proliferation across all treatments. Similarly, a trend was observed where high responders displayed greater nitric oxide production compared to average and low responders across heat treatments.

**Conclusion:** Overall, these results suggest that blood mononuclear cells from high immune responder dairy cows are more thermotolerant compared to average and low immune responders

**Background**

Increases in greenhouse gases, over the years, are contributing to climate change. This is because greenhouse gases trap energy, that is emitted from the earth's surface, in the lower atmosphere therefore not allowing this energy to be emitted back into space [1]. These trapped greenhouse gases cause an overall increase in global temperatures, termed global warming. Global warming causes an increase in the overall number of hot days and heat waves, and a decrease in the number of cooler days or cooling at night [1]. Global warming directly impacts livestock welfare by increasing incidence of heat stress. Dairy cows have been identified as the livestock species most sensitive to elevated temperatures and humidity due to their high metabolic heat load [2] and therefore are quite susceptible to heat stress.

Heat stress negatively affects the health, welfare, and production of dairy cattle leading to increased disease incidence [3] and impairing immune function [4]. Studies have shown heat stress causes an increase in the incidence of several different diseases in dairy cattle, including metabolic disorders, as well as mastitis[5, 6]. Similarly, increased mortality in dairy cattle has also been reported as a result of heat stress [7]. These health-related issues could be due to the impaired immune function observed in heat stressed dairy cattle. Heat stress can impact both the innate and adaptive arms of the immune system. One of the effects of heat stress on the adaptive immune response is disruption of the balance between T-helper 1 (TH1) and T-helper 2 (TH2) responses, causing a shift towards a TH2 response [4]. This bias can lead to impaired cell-mediated immune response (CMIR). Similarly, heat stress in dairy cattle has been shown to reduce lymphocyte proliferation [8, 9]. Lymphocytes, which include B and T-cells, rapidly proliferate in response to invading pathogens in order to facilitate clearance. Accordingly, if lymphocyte proliferation is reduced it is more difficult for cattle to defend against invading pathogens.

Another important defence mechanism of the immune system is the production of molecules that destroy or damage pathogens. Nitric oxide is one such molecule that plays a key role in defense against pathogens [10]. Additionally, nitric oxide has been identified as a molecule that also plays a role in thermotolerance. Studies in ruminant species have shown nitric oxide production as one of the key markers associated with thermotolerance [11, 12]. This is likely because nitric oxide appears to play a crucial role in the vasodilation of blood vessels of the skin during heat stress [13] Therefore, not only does nitric oxide play a role in protection against pathogens it may also play a significant role in protecting ruminants, like dairy cattle, against the negative impacts of heat stress.

The primary cellular protection mechanism against heat stress is the production of heat shock proteins (HSP), with HSP70 being the most abundant HSP produced during the heat shock response. Even a slight increase in core body temperature can cause misfolding of cellular proteins leading to the disruption of cell organelle organization and impaired intercellular transport processes [14]. Ultimately if temperature can not be regulated and corrected the cell will die [14]. Therefore, HSP expression increases during periods of elevated environmental temperatures [14]. Heat shock proteins protect cells from damage caused by heat stress. They assist in repairing the cell and facilitate the re-folding of proteins to return the cell to its original state [14–16]. Therefore, the production of HSP is critical for protection against cellular damage caused by heat stress.

Previous studies have shown the ability to identify and selectively breed animals that have enhanced immune response, based on estimated breeding values (EBVs) [17, 18]. Dairy cattle identified as high immune responders based on their EBVs demonstrate several health benefits; for example, studies in lactating dairy cattle have shown cattle with enhanced antibody-mediated immune response (AMIR) and/or enhanced CMIR have reduced occurrence of a number of common diseases including mastitis, ketosis, metritis, displaced abomasum and retained fetal membrane [19–21]. Similarly, calves identified as high antibody responders have been shown to have reduced occurrence of diarrhoea [22]. Not only do high immune responding cattle have a reduced occurrence of disease, they are also more responsive to commercial vaccines compared to animals classified as low immune responders [23]. More recently it has also been shown that high immune response cattle have better hoof health in terms of lower incidence of digital dermatitis [24]. Additionally, high immune responder dams also provide benefits to their calves by passing on their high immune response genetics and also by providing colostrum that contains increased total immunoglobulin-G and β-lactoglobulin [25].

To date the majority of these studies indicating the benefits of being a high responder, have been conducted in purebred Holstein cattle. However, previous work has compared immune response traits between purebred Holsteins and crossbred dairy cattle. For example, studies in calves indicated crossbred calves had significantly greater antibody response and a trend towards greater survival to first calving than purebred Holstein calves [26]. Similarly, it was also observed that crossbred first calf heifers had both significantly greater AMIR and CMIR compared to purebred Holsteins [27]. Although the results on immune response in other breeds compared to Holsteins are encouraging, the Holstein breed remains the dominant dairy breed due to its greater ability to milk than other dairy breeds. Therefore, due to higher metabolic heat production, Holstein cattle may be more prone to heat stress [28]. Since, health and welfare benefits have been observed in high immune responding Holstein cattle, a breed more prone to heat stress than others, it was evaluating blood mononuclear cells (BMC) from immune-phenotyped Holstein cattle for cellular responses and indicators of thermotolerance. Therefore, the objective of this study was to evaluate the effect of *in-vitro* heat challenge on BMC of Holstein cows ranked as high, average, and low immune responders, for HSP70 concentration, nitric oxide production and cell proliferation.

**Results**

*AMIR and CMIR*

Over the last 5 years (2014-2019), 1056 Holstein dairy cows from the University of Guelph Elora Research Dairy Herd have been evaluated based on EBVs for their AMIR and CMIR phenotype. It was observed that a total of 182 cattle were classified as high AMIR, 681 were classified as average AMIR and 193 were classified as low AMIR. Similarly, a total of 160 were classified as high CMIR, 726 classified as average CMIR and 170 classified as low CMIR.

For the purposes of this study 15 animals were selected from the high AMIR-high CMIR phenotype, 15 from the Low AMIR-Low CMIR phenotype and 15 from the average AMIR-average CMIR phenotype. A total of 45 animals were selected to evaluate the effects of heat stress on BMC function.

*HSP70 Concentration*

Across treatments, high immune responders had a significantly higher HSP70 concentration after Heat Stress 1 (HS1, one heat challenge at 42°C) (10.58ng/ml) compared to thermoneutral (TN) (9.53ng/ml) (p = 0.03), and a significantly lower HSP70 concentration 18-hour post recovery after HS1 (8.83ng/ml) compared to HS1 (10.58ng/ml) (p = 0.03). After the second heat challenge (HS2, two subsequent heat challenges at 42°C), HSP70 concentration increased, in cells from high immune responders, compared to the HSP70 concentration after the 18-hour recovery from HS1, but not significantly. No significant differences were observed for HSP70 concentration between heat treatments for the average responders, except for 18-hours after HS2 (10.21ng/ml), which was significantly greater than at HS2 (9.06ng/ml, p=0.01) and 18-hour after HS1 (8.63ng/ml, p=0.03). Low immune responders had significantly greater HSP70 concentrations at HS1 (9.85ng/ml) compared to TN (8.46ng/ml, p=0.006). However, HSP70 concentration decreased to its lowest concentration after HS2 (8.45ng/ml), with both HS1 (p=0.02) and 18-hours post HS1 (9.54ng/ml, p=0.03) having significantly greater HSP70 concentration compared to HS2.

Results in figure 1 show a trend (p=0.07) for BMC from high responders to start (TN) with a higher HSP70 concentration compared to BMC from low responders. Cells from high and low immune responders also had significantly greater HSP70 concentrations compared to BMC from average responders after the first heat challenge (HS1). No significant differences were observed between cells from all phenotypes 18-hours after HS1. However, after the second heat challenge (HS2) there was a trend (p=0.06) for BMC from high responders to have greater HSP70 concentration compared to BMC from low responders. Similarly, BMC from high responders had significantly greater HSP70 concentrations 18-hours after HS2 compared to BMC from low responders. Cells from average responders trended towards greater HSP70 concentration compared to BMC from low responders 18-hours after HS2.

*Nitric Oxide Production*

Significant differences in nitric oxide production between unstimulated and stimulated cells within each treatment were observed (TN p = 0.007, HS1 p = 0.03, HS2 p = 0.04). The general trend was for nitric oxide production to decrease between each heat treatment, with stimulated TN BMC producing significantly greater nitric oxide than stimulated cells subjected to two heat stress challenges (HS2, p = 0.009). Although not significant, stimulated TN cells also produced more nitric oxide than stimulated HS1 cells.

When comparing within phenotype across treatments, the general observation was for BMC from high responders to decrease nitric oxide production between treatments. Stimulated cells from high responders produced significantly more nitric oxide under TN conditions than after HS2 (p= 0.03). Similarly, although not significant, stimulated cells from high responders produced more nitric oxide under TN conditions than after HS1. No significant differences were observed in nitric oxide production across all heat treatments for stimulated or unstimulated cells from average responders. Similar to what was observed in cells from high responders, the general pattern was for cells from low responders to decrease nitric oxide production across all heat treatments. In low responders a trend was observed for stimulated TN cells to produce more nitric oxide than stimulated cells after HS2 (p = 0.09). Although not significant, stimulated TN cells from low responders produced more nitric oxide than stimulated HS1 cells.

Results shown in Figure 2 indicate that BMC from high immune responders produced a greater concentration of nitric oxide across all treatments compared to cells from average and low responders (p = 0.04). Similarly, stimulated BMC from high responders produced significantly greater (0=0.007) concentrations of nitric oxide under TN conditions compared to TN stimulated cells from average responders. There was a trend (p=0.09) for cells from high responders to produce greater concentrations of nitric oxide when unstimulated in TN conditions compared to unstimulated TN cells from average responders. No significant differences were observed within other treatments between phenotypes.

*Cell Proliferation*

The overall trend observed in cell proliferation assays was for cells from high immune responders to increase proliferation after each heat treatment. Results showed significantly greater cell proliferation at HS1 (2.81, p = 0.04) and HS2 (3.24, p=0.01) compared to TN (1.65). Although not significant, cell proliferation increased after HS2 compared to HS1. Conversely, the overall observation for cells from low responders was for cell proliferation to decrease between heat stress treatments. Cell proliferation at TN was significantly greater than cell proliferation after HS2 (p= 0.001) for low responders and although not significant, cell proliferation at TN was higher than cell proliferation after HS1. Similarly, cell proliferation after HS1 was significantly greater than cell proliferation after HS2 (p= 0.03) for low responders. No significant differences were observed for cell proliferation between any of the heat treatments for average responders.

Results in Figure 3 indicate that BMC from high immune responders trended (p = 0.08) towards greater cell proliferation at TN compared to low responders. No significant differences were observed for the TN treatment between average and high, or between average and low immune responders. It was observed that after the first heat challenge (HS1), cells from high immune responders underwent significantly greater proliferation compared to cell from both average (p = 0.04) and low (p = 0.02) responders. However, no significant difference was observed between average and low responders for cell proliferation after one heat challenge (HS1). Similarly, BMC from high responders displayed significantly greater proliferation compared to cells from both average (p = 0.04) and low (p = 0.03) responders upon subsequent heat challenge (HS2). Average immune responders also trended (p = 0.05) towards greater cell proliferation compared to low responders after the second heat challenge (HS2).

**Discussion**

Heat stress negatively impacts cells, including cells of the immune system. Under heat stress cellular proteins become denatured or misfolded which may transient cell cycles, or create changes in cell structure, all of which could eventually lead to cell death [29–31]. Heat stress induces the rapid initiation of the heat shock response, which involves increased expression of heat shock proteins, especially HSP70 [15, 32, 33]. Results from the current study showed that BMC from cattle identified as high immune responders based on their AMIR and CMIR-EBVs produced significantly greater concentrations of HSP70 after on heat stress event (HS1) compared to TN. Studies have shown HSP70 is produced in response to heat stress and acts intracellularly in order to prevent the accumulation of mis-folded proteins or re-fold proteins that have been denatured [34]. Therefore, the results described here indicate that BMC from high immune responder Holstein cattle are able to effectively respond to an initial *in-vitro* heat stress challenge by producing increased HSP70 that could act to repair and protect cells during heat stress. Additionally, BMC from high immune responders produced significantly greater concentrations of HSP70 after HS1 compared to 18hrs after HS1. It has been reported that once heat stress has subsided the production of HSP70 declines, with it being suggested that this occurs by HSP70 binding to and degrading its own mRNA in an effort to stop transcription [32]. These results suggest that by 18hrs after HS1 BMC from high immune responders were able to counteract the impact of heat stress at the cellular level and return HSP70 concentrations to similar concentrations observed at TN.

Upon a secondary exposure to heat challenge although HSP70 concentration did increase slightly in BMC from high responders, significant differences compared to TN were not observed until 18hrs after HS2. It is possible that multiple exposures to heat may have impacted the function of the cells resulting in delayed production of HSP70 concentrations that are significantly different than TN. As discussed previously, heat stress can cause morphological changes and alterations in cell cycles which can affect cellular function [30, 31]. The impact of heat stress on cellular function may vary among cattle depending on the level of thermotolerance [9].

Similar results regarding HSP70 production, to those discussed of high immune responders, were also observed in BMC from low immune responders. Cells from low immune responders produced greater concentrations of HSP70 after HS1 compared to at TN. However, in contrast upon second exposure to heat challenge (HS2) no significant difference was observed in BMC from low responders for HSP70 production after HS2 and 18hrs after HS2 compared to at TN. In fact, the HSP70 concentration at HS2, in cells from low responders, was significantly less than the concentration after HS1 and 18hrs after HS1. These results may indicate that upon initial exposure to a heat challenge low responders are able to effectively respond by producing HSP70. However, upon multiple exposures to heat challenge the cells of low responders may become damaged and possibly even start to die resulting in concentrations of HSP70 similar to that observed under TN conditions. Indeed, previous studies have shown a link between HSP production and cell proliferation both in non-stress and stress conditions [35]. This is supported by results from cell proliferation experiments (discussed later) in low responders since it was observed that BMC from low responders displayed significantly reduced proliferation after HS2 compared to HS1 and TN.

Conversely, no significant difference in HSP70 concentration, for any of the heat challenge treatments compared to TN, was observed for BMC from average responders. These results were not expected, however they do correspond those for cell proliferation in cells from average responders, where no difference was observed between treatments. Therefore, the lack of a difference in HSP70 concentration between treatments, in BMC from average cows, could in part explain why no difference was observed in cell proliferation as well. Additionally, heat stress can also impair the CMIR, by shifting towards a TH2 response [4]. Therefore, it is possible that the average responders, sampled in this study, had a shift more towards a TH2 biases resulting in impaired cellular responses.

Results comparing the concentration of HSP70 in BMC from high, average and low responders, indicated cells from that high immune responders produced significantly greater concentrations of HSP70 compared to cells from average immune responders after the first heat challenge (HS1), and a trend towards greater HSP70 production compared to low immune responders after two heat challenges (HS2). These results suggest that dairy cattle identified as high immune responders have an enhanced ability to protect and repair cells during both a single heat stress event and multiple heat stress events, conditions similar to that of a heat wave [34, 36, 37]. Additionally, elevated production of HSP70 from BMC of high immune responder dairy cows may indicate that high immune responder cattle are more thermotolerant than average and low herd mates. Indeed, previous studies in cattle have shown an association between increased expression of HSP70 and improved thermotolerance [38]. Several genome-wide association studies performed with thermotolerant cattle, show various genes significantly associated with heat shock protein and immune response [11, 28]. These studies indicated that HSP70 and immune response genes play key roles in thermotolerance in dairy cattle. Therefore, this study may provide evidence that cattle with enhanced immune response may be more thermotolerant compared to their herd mates.

Nitric oxide synthesis one such immune response trait that has been identified as a mediator of in thermotolerance in ruminant species [11]. Specifically, nitric oxide is required to facilitate vasodilation of ruminant's skin during hyperthermia. In the current study it was observed that BMC from high immune responder dairy cows tended to produce more nitric oxide across all treatments compared to average and low responders. This could indicate an enhanced ability, of high responders, to dissipate heat during heat stress compared to average and low responder herd mates, potentially indicating an improved ability to thermoregulate during a heat event. Upon sensing increases in body temperature, nitric oxide is produced by BMC. This activates cutaneous vasodilation, with enhanced production of nitric oxide being essential for increased vasodilation of skin during heat stress [12].Vasodilation of skin is one mechanism ruminants use to dissipate heat during heat stress [5]. Therefore, the higher production of nitric oxide in cells from high responders provides that evidence high responders may have greater vasodilation, compared to average and low responders, potentially leading to improved thermotolerance.

Not only is nitric oxide associated with thermoregulation, it is also an important molecule produced in response to pathogenic infections. Nitric oxide neutralizes pathogens through oxidative damage. Oxidative stress alters microbial DNA, inhibits enzyme function, and induces lipid peroxidation [10], mechanisms which contribute to the elimination of a pathogen from its host. Blood mononuclear cells from high immune responder cattle, stimulated by lipopolysaccharide (LPS) produced more nitric oxide than cells from average and low responders under all test conditions. This may indicate that high immune responder cattle have a better ability to defend against pathogens during heat stress compared to average and low responding cattle.

Another negative impact of heat stress is decreased lymphocyte proliferation [8] and impaired cellular function [9], factors which may impair the ability of the immune system to respond to an invading pathogen by reducing the magnitude of response generated to the pathogen. Results from this study showed that BMC from dairy cattle identified as high immune responders underwent greater cell proliferation when stimulated with the mitogen Concanavalin A (ConA) compared to average and low responders after two heat treatments. Greater cell proliferation after heat treatment provides evidence that high immune responder cattle may have an increased capacity to respond to infection during heat stress, compared to average and low herd mates. Consequently, high responders may experience lower incidence of disease and mortality during times of heat stress compared to average and low responders.

Overall, cell proliferation in high immune responders increased across treatments, whereas cell proliferation remained the same for average cows, and decreased for low responder cows. As mentioned previously heat stress is typically associated with reduced lymphocyte proliferation. Additionally, when referring to the stress response in general, acute stress typically causes the immune system to be stimulated and poised for response to pathogens, whereas chronic stress typically leads to immune suppression and programmed cell death of immature lymphocytes [39]. Therefore, these results may indicate the *in vitro* heat challenges imposed in this study resulted in an acute stress for high responders and potentially average responders, however resulted in a more chronic stress for low responders. As previously mentioned, other studies showed the production of HSP being associated with cell proliferation, however the direction of the effect can vary For example, some studies show an association between increased cell proliferation and production of HSP during certain conditions of stress (likely acute stress), where it is suggested that the increased production of HSP may be involved in inhibiting the function of proteins that degrade proliferative-specific proteins [35]. Additionally, other studies indicated reduced cell proliferation has also been associated with HSP production during heat stress, which is likely the result of cell death due to chronic stress conditions [40]. Based on the results of this study from *in vitro* HSP70 and nitric oxide production and BMC proliferation assays, there is evidence that BMC from high immune response Holstein dairy cattle appear to have a better ability to tolerate exposure to increased heat and respond better to heat stress compared to BMC from low immune response herd mates. To confirm these results of this study in-vivo studies evaluating similar traits could be performed, where physiological parameters are measured as well.

**Conclusion**

In conclusion results from this study provide evidence that BMC from Canadian Holstein dairy cows identified as having enhanced or high immune response based on their AMIR and CMIR-EBV may be more thermotolerant compared to average and low immune responders. This is based on the discovery that cells from high immune responder cattle exhibited enhanced production of HSP70, an important molecule for the protection of cells during heat stress and one of the key molecules involved in thermotolerance. Moreover, BMC from high immune responders trended towards increased nitric oxide production, another molecule identified as having a role in thermotolerance. Additionally, it was observed that BMC from high responders displayed increased proliferation, when stimulated by the mitogen ConA, and enhanced nitric oxide production, when stimulated with LPS, following *in vitro* heat stress. Therefore, high immune responder Holsteins, already reported to have enhanced resistance to disease, may also have improved immunity during periods of heat stress. Selecting for animals with enhanced immune response genetics may improve overall resilience, not only by improving immunity, but by increasing resistance to heat stress and improving thermotolerance.

**Methods**

*Animals*

The dairy cattle used in this study were housed at the University of Guelph Elora Research Station. All cattle were housed in the lactation wing of the barn, a large free-stall area with sufficient bunk space for each animal. Cattle are cared for by the staff at the research station, with the facility being required to follow guidelines set out by the University of Guelph Animal Care Committee. Additionally, all lactating cattle at this facility were previously phenotyped for immune response. Cattle were evaluated for immune response using a patented test protocol described by Wagter et al. (2000) and Hernandez et al. (2005) with some modifications as described here. Briefly, blood samples were taken prior to immunization on Day 0 to assess baseline specific antibody. Cattle were then immunized intramuscularly with type 1 and type 2 antigens (US Patent #7,258,858; Wagter and Mallard, 2007). On Day 14, another blood sample was taken to assess AMIR. A cutaneous delayed type hypersensitivity (DTH) test was also initiated on day 14 to assess CMIR. To do this, initial measurements were taken of the tail-fold thickness in triplicate. A phosphate saline buffer (PBS) was injected intra-dermally into the left tail fold as a control and the type 1 test antigen was injected intra-dermally into the right side. Twenty-four hours later final skin fold measurements were taken to assess DTH as an indicator of CMIR. The blood samples taken on Day 0 and Day 14 were spun at 700xg for 15 minutes and sera was obtained. The sera were stored at -20°C until time of analysis. Sera samples were used to evaluate specific antibody response to the type-2 antigen by indirect enzyme-linked immunosorbent assay (ELISA) that was previously developed in the lab [43]. The cellular response was determined by averaging the triplicate measurements taken on Day 14 and 15 for control and test sides. A ratio of (test 24 – test 0)/ control 24 – control 0) was used to assess CMIR. Pedigree information was obtained for all animals (n=1056) and EBVs were calculated for AMIR and CMIR using ASREML, considering the effects of age, pregnancy status, and year of testing. Cattle were ranked as high for AMIR or CMIR if their EBV was one standard deviation above the population mean and were ranked as low if their EBV was one standard deviation below the population mean. From the 1056 animals that were ranked for immune response, a total of 45 lactating Holsteins, of varying parity status, in all stages of lactation and pregnancy status, were selected for this study. Of these 45 animals, 15 were ranked as high for AMIR and CMIR, 15 ranked as average, and 15 ranked as low for both traits (Sample calculation was done using one-way ANOVA test in G*Power 3.1.9.7 [44], using α of 0.05, power of 0.95 and standard deviation and means from previous unpublished data). Average AMIR average CMIR phenotyped animals for this study had EBVs between -0.2 and 0.2 to ensure the EBV for the average AMIR average CMIR animals were as close to zero as possible. Since the objective of the study was to compare BMC function, between high, average, and low immune responders, after exposure to heat challenge no control group of animals was used in this study. The 45 cattle were bled in the winter months in order to avoid previous exposure to heat. Whole blood samples were obtained via the tail vein from each animal. Animals were bled in groups of 6, with 2 high, 2 average and 2 low being bled in each group. Groups were assigned at the outset of the study in order to keep the phenotype of each animal anonymous during the data collection period. All data collected on each animal in the study was used to assess the function of BMC in high, average, and low immune responders. Upon completion of the study, all animals were permitted to remain in the herd at the Elora Dairy Research Station and are available for use in other subsequent studies.

*BMC Isolation*

Whole blood was centrifuged at 1200xg for 20 minutes with the centrifuge break turned off. The buffy coat was isolated and diluted in a 1:2 ratio with PBS. The buffy coat PBS solution was then layered over histopaque (Sigma-Aldrich, Oakville, ON.) in a 1:1 ratio. The histopaque buffy coat solution over lay was then centrifuged at 1200xg for 10 minutes with the break off. The buffy coat was isolated, and PBS was added to a total volume of 25ml. The cell suspension was then centrifuged at 150xg for 10 minutes with the brake on. The supernatant was discarded, and the cell pellet was resuspended in 3 ml of PBS. To lyse red blood cells, present in the cell pellet, the cell pellet was resuspended in 1ml of deionized water (Thermo-Fisher Scientific, Ottawa, ON.) and PBS was immerdiately added to a total volume of 25 ml. The cell suspension was again centrifuged at 150xg for 10 minutes with the brake on. The supernatant was discarded, and the cell pellet was resuspended in 3 ml of PBS. The total number of cells per ml was determined using an ORFLO cell counter (ORFLO Technologies, Ketchum, ID). Cell viability was checked using trypan blue solution (Sigma-Aldrich, Oakville, ON.) via hemocytometer method.

*Heat Challenge*

The heat challenge protocol conducted in this study was done in order to mimic what might occur during a heat wave. Viable cells were plated in the concentrations described below for each assay. All plates were placed in a 37°C incubator overnight (18 hrs). On the subsequent day plates being exposed to heat challenge were removed from the 37°C incubator and placed in a 42°C incubator for 4 hours (heat challenge 1(HS1) and then all plates except HSP- HS1 were returned to the 37°C incubator after the first heat challenge was completed. Supernatant and cells from HSP-TN (to represent normal environmental conditions; 22 hrs at 37°C) and HSP- HS1 (to represent the initial stages of a heat wave 18 hrs at 37°C + 4 hrs at 42°C, 22 hrs total) plates were collected, following the protocol described below, immediately following completion of HS1. The day following HS1 (the first heat challenge), plates undergoing a second heat challenge (in order to represent a full heat wave (HS2)) were removed from the 37°C incubator and placed in a 42°C incubator for 4 hours. Additionally, 18hr HSP-HS1 plates (plates subjected to one heat challenge the previous day and to represent the brief recovery period for HSP70 after the initial stage of a heat wave 18hrs at 37°C + 4hrs at 42°C + 18hrs at 37°C, 40hrs total) were removed from the 37°C incubator, 18 hrs after the HS1 heat challenge, and supernatant and cells were collected as described below. After HS2 was completed all plates except HSP-HS2 plates were removed and placed back in the 37°C incubator until time of analysis or sample collection depending on the assay conducting. Supernatant and cells were collected, as described below, from HSP-HS2 plates (18hrs at 37°C + 4hrs at 42°C + 18hrs 37°C + 4hrs 42°C, 44 hrs total) immediately following the HS2 heat challenge. Supernatant and cells from 18hr HSP-HS2 plates (plates that represent recovery period for HSP70 after full heat wave 18hrs at 37°C + 4hrs at 42°C + 18hrs at 37°C + 4hrs at 42°C + 18hrs at 37°C, 62hrs total) were collected, as described below, 18 hrs after the completion of the HS2 heat challenge. Thermoneutral plates for nitric oxide production and cell proliferation remained in the 37°C incubator the entire duration of the heat challenge. A summary of the heat challenge and incubation steps for each assay is presented in table 1.

Table 1: Summary of treatments for evaluating HSP70 concentration, nitric oxide production and cell proliferation.

|  | Assay Type | | |
|---|---|---|---|
| **Heat Treatment** | HSP70 Production | NO Production | Cell Proliferation |
| TN (37°C) | Plate remains in 37°C until immediately after HS1 (22hrs 37°C) (collect @ same time as HS1 plate) | Plate remain at 37°C until 48hrs after LPS stimulation | Plate remain at 37°C until 72hrs after ConA stimulation |
| HS1(1x 42°C for 4hrs) | All plates, but TN, removed from 37°C after overnight incubation and placed at 42°C. Samples collected immediately after 4hrs 18hrs 37°C + 4hrs 42°C). Other plates returned to 37°C | All plates, but TN removed from 37°C after overnight incubation and placed at 42°C for 4hrs, then returned to 37°C | All plates, but TN removed from 37°C after overnight incubation and placed at 42°C for 4hrs, then returned to 37°C |
| 18-hrs HS1 | Plate removed from 37°C 18hrs after HS1 treatment and samples collected (18hrs 37°C + 4hrs 42°C + 18hrs 37°C) | Not applicable | Not applicable |
| HS2(2x 42°C for 4hrs) | Remaining 2 plates removed from 37°C and placed at 42°C. Samples collected from HS2 plate immediately after 4hrs (18hrs 37°C + 4hrs 42°C + 18hrs 37°C + 4hrs 42°C). 18-hrs HS2 plate returned to 37°C | Only HS2 plates removed from 37°C, following another overnight incubation, and placed at 42°C for 4hrs. Then return to 37°C until 48hrs after LPS stimulation when samples from all plates collected | Only HS2 plates removed from 37°C, following another overnight incubation, and placed at 42°C for 4hrs. Then return to 37°C until 72hrs after Con A stimulation when MTT assay for all plates performed. |

| 18-hrs HS2 | Plate removed from 37°C 18hrs after HS2 treatment and samples collected (18hrs 37°C + 4hrs 42°C + 18hrs 37°C + 4hrs 42°C + 18hrs 37°C) | Not applicable | Not applicable |

*HSP70 Concentration*

The cell suspension was diluted in prewarmed RPMI media (Thermo-Fisher Scientific, Ottawa, ON.) containing 10% heat inactivated fetal bovine serum (Sigma Aldrich, Oakville, ON.) and Penicillin-Streptomycin (Sigma Aldrich, Oakville, ON) to a concentration of $1.0 \times 10^6$ cells/ml. Cells were then plated in three 1ml replicates in 24-well flat bottom cell culture plates (Sigma Aldrich, Oakville, ON.) and placed in a 37°C with 5% $CO_2$ incubator overnight. On the following day all plates except for the TN plates were moved to an incubator that was set at 42°C with 5% $CO_2$ for 4 hours and then returned to the 37°C incubator as recommended in previous heat stress studies [9, 45]. The cell culture supernatant and the cells were collected from each well of the TN plates and the plates that were subjected to one heat treatment (HS1, 4-hour heat challenge in 42°C incubator). To collect the cells and the supernatant, the cell supernatant was collected from each well and the 3 wells for each cow were pooled. Additionally, 200 µl of Trypsin, with a concentration of 5000 mg/L, (Thermo-Fisher Scientific, Ottawa, ON.) was added to each well and placed back in the 37°C incubator for 5 minutes to remove any cells that were stuck to the plate. The Trypsin suspension for each animal was added to the pooled cell culture supernatant and the whole suspension was centrifuged at 2500xg for 10 minutes at 4°C. The supernatant was discarded, and cells were lysed using a combination of Halt protease inhibitor (Sigma Aldrich, Oakville, ON.) at a concentration of 10µl/ml and M-PER Mammalian

Protein Extraction Reagent (Thermo-Fisher Scientific, Ottawa, ON.). In order to remove cell debris, the lysed cell suspension was centrifuged at 2500xg for 30 minutes at 4°C. The liquid portion containing the protein was then aliquoted and stored at -80°C until time of analysis. Eighteen hours after the HS1 treatment, cells were obtained and processed similarly to what was described above, in order to assess recovery HSP70 concentration after one heat challenge (18hrs HS1). The remaining plates were again placed into the 42°C incubator for 4 hours to assess HSP70 concentration after 2 heat challenges on subsequent days (HS2). This is to mimic diurnal and cooler nocturnal heat patterns. Cells from the one plate were obtained and processed as described above, whereas the other plate was placed back in the 37°C incubator overnight. Eighteen hours following the HS2 treatment cells were obtained and processed from the final plate, as described above, in order to assess recovery HSP70 concentration after two subsequent heat challenges (18hrs HS2). All the samples were assessed for HSP70 concentration using a commercial ELISA kit (Abclonal, Woburn, MA) following manufactures instructions. Extra protein fraction was obtained from one of the animals and used on all ELISA plates as an internal control to ensure concentrations could be compared across all plates.

*Nitric Oxide Production*

The cell suspension was diluted to a concentration of $1.0 \times 10^6$ cells/ml in Gibco serum free cell-culture media (Thermo-Fisher Scientific, Ottawa, ON.) containing sodium pyruvate (Thermo-Fisher Scientific, Ottawa, ON.) and MEM amino acids solution (Thermo-Fisher Scientific, Ottawa, ON.). The cells were plated in six 1ml replicates per animal on 24 well cell culture plates (Sigma-Aldrich, Oakville, ON.) and placed in a 37°C incubator overnight. The following day LPS, from Escherichia coli O127:B8 (Sigma-Aldrich, Oakville, ON), was added to 3 of the replicates from each animal in a concentration of 5µg/ml. The cells were exposed to LPS in order to observe the effects of pathogen exposure during heat stress. The TN treatment plates were returned to the 37°C incubator whereas the HS1 and HS2 plates were transferred to a 42°C incubator for 4 hrs. After the 4-hour heat treatment all plates were returned to a 37°C incubator. The following day the HS2 plates were again transferred to a 42°C incubator for 4 hrs and then returned to a 37°C incubator. All plates remained in a 37°C incubator until 48 hrs after stimulation with LPS. At this time all plates were removed from the incubator and the culture supernatant was collected from all wells. The replicates of either stimulated or unstimulated cells from each animal were pooled within their respective treatments and samples were stored at -80 °C until time of analysis. Nitric oxide production was determined by Griess Assay (Promega, Madison, WI) following the manufacture's instructions. Nitric Oxide production was determined using a standard curve following instructions provided by the manufacturer.

*Cell Proliferation*

The cell suspension was diluted in RPMI media (Thermo-Fisher Scientific, Ottawa, ON.) containing fetal bovine serum (Sigma Aldrich, Oakville, ON) and Penicillin-Streptomycin (Sigma Aldrich, Oakville, ON) to concentration of $5.0 \times 10^5$ cells/well. Cells were plated in 96-well flat bottom culture plate (Sigma Aldrich, Oakville, ON) in six 100 µl replicates. Six replicates of culture media were also added to each plate as a control. All plates were placed in a 37°C incubator overnight. The following day ConA (Sigma Aldrich, Oakville, ON) was added in a concentration of 5µg/ml to 3 of the 6 replicates for each animal on all plates. Two of the plates were then transferred to a 42°C incubator for 4 hrs and the remaining plate was returned to the 37°C incubator for the TN treatment. After the 4-hour heat treatment the 2 plates were returned to the 37°C incubator. The following day, one of the plates that was subjected to HS1 on the previous day was placed back in the 42°C incubator for 4 hrs, for the HS2 treatment, and then placed back in the 37°C incubator. All plates remained in the 37°C incubator until 72 hrs after ConA stimulation. At this time cell proliferation was evaluated for all treatments using the MTT assay (Sigma Aldrich, Oakville, ON.) following the manufacturer's instructions with minor modifications. After MTT solution was added and the cells were incubated for 3 hrs at 37°C, plates were centrifuged at 600xg for 5 minutes. Culture supernatant was then discarded and MTT solvent was added to wells. Plates were placed on a gyratory shaker for 10 mins and read at a wavelength of 570nm. All replicates for each animal were averaged with the coefficient of variation for each sample being 10% or less. Cell proliferation was determined by calculating the ratio based on the following formula: (simulated cells − culture media)∕(unstimulated cells − culture media).

*Statistical Analysis*

Data from all assays were checked for normality using the Shapiro-Wilks test in R 3.6 (R Core Team 2019). Any raw data that were not normally distributed was log transformed and re-checked for normality using the Shapiro-Wilks test. All data were analyzed using general linear models in R 3.6 [46].

To assess differences across treatments within each assay the following model was used:

$$y_{ijklm} = \mu + l_i + s_j + p_k + i_l + t_m + e_{ijklm}$$

where $y_{ijklm}$ = the results from particular assay (ie cell proliferation, nitric oxide production, HSP70 concentration), $\mu$ = overall mean, $l_i$ = effect of lactation number ( lactation 1, 2, 3 and ≥4), $s_j$= effect of stage of lactation (1 = 1-100 days in milk (DIM), 2 = 101-200 DIM, 3 = 201-305 DIM), $p_k$ = effect of pregnancy status (0 = not pregnant, 1= 1-100, 2= 101-200, 3= >200 days in calf), $i_l$ = effect of immune response phenotype (high, average or low), $t_m$ = the effect of heat treatment (TN, HS1 or HS2 for cell proliferation and nitric oxide and TN, HS1, 18hrs HS1, HS2, 18hrs HS2 for HSP70 concentration) and $e_{ijkl\text{m}}$= residual error.

The following model was used to assess differences between phenotype for each assay, within each heat treatment:

$$y_{ijkl} = \mu + l_i + s_j + p_k + i_l + e_{ijkl}$$

where $y_{ijkl}$ = the particular heat treatment within each assay (ie TN, HS1 or HS2 for cell proliferation and nitric oxide and TN, HS1, 18hrs HS1, HS2, 18hrs HS2 for HSP70 concentration), $\mu$ = overall mean, $l_i$ = effect of lactation number ( lactation 1, 2, 3 and ≥4), $s_j$= effect of stage of lactation (1 = 1-100 days in milk (DIM), 2 = 101-200 DIM, 3 = 201-305 DIM), $p_k$ = effect of pregnancy status (0 = not pregnant, 1= 1-100, 2= 101-200, 3= >200 days in calf), $i_l$ = effect of immune response phenotype (high, average or low) and $e_{ijkl}$ = residual error.

All results are presented as least squared means of untransformed data; however, p-values are based on normalized data. Significance is reported at $P < 0.05$, whereas trends are reported at $P < 0.10$.

**List of abbreviations**

**AMIR:** antibody-mediated immune response

**BMC:** blood mononuclear cells

**CMIR:** cell-mediated immune response

**ConA:** Concanavalin A

**DIM:** days in milk

**DTH:** delayed type hypersensistivity

**EBV:** estimated breeding value

**ELISA:** enzyme linked immunosorbant assay

**HS1:** heat stress 1 (1$^{st}$ heat challenge)

**HS2:** heat stress 2 (2$^{nd}$ heat challenge)

**HSP:** heat shock protein

**LPS:** lipopolysacchride

**PBS:** phosphate saline buffer

**TH1:** T-helper 1

**TH2:** T-helper 2

**TN:** Thermoneutral

**Figures**

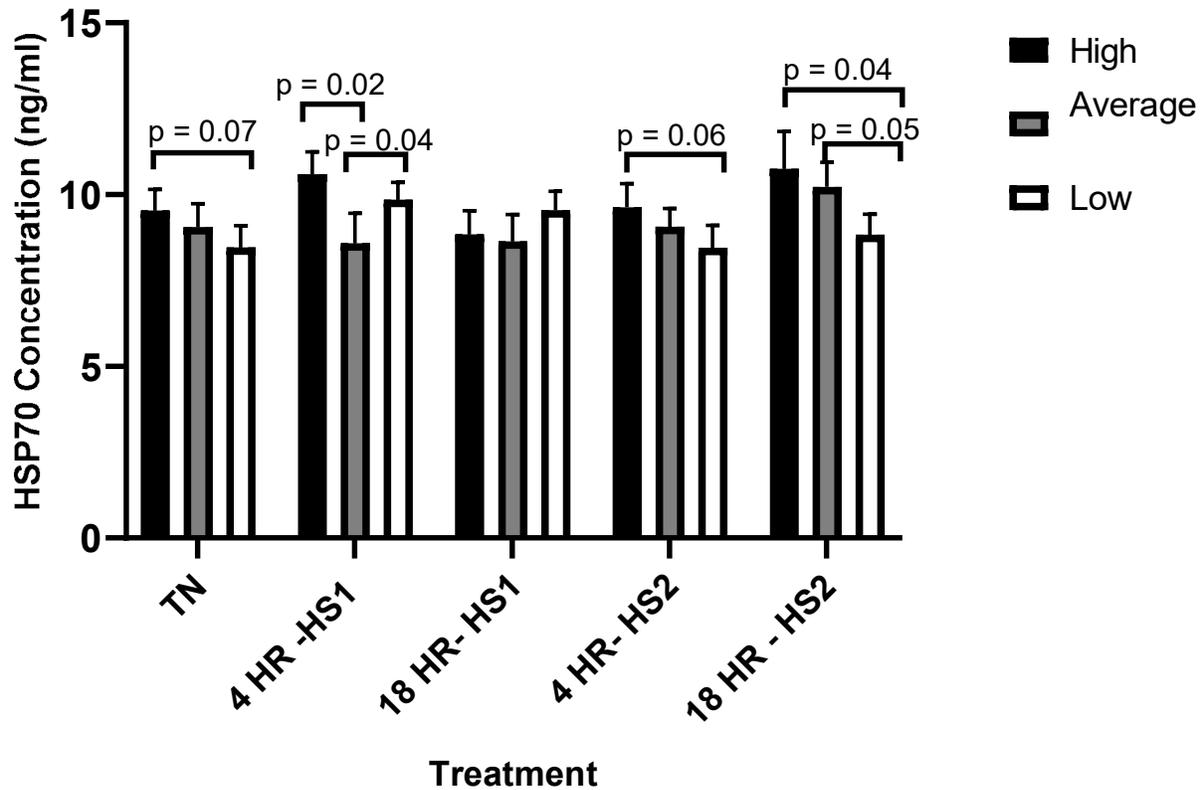

**Figure 1: The concentration of heat shock protein 70, by heat treatment, for high, average and low immune responders.** Cells were subjected to various heat treatments and heat shock protein 70 concentration was measured by ELISA. The data presented are least squared means + standard error of the mean. High immune responders are represented by the black bars, average immune responders by the grey bars and low immune responders by the white bars. P-values for significant effects or trends, between immune response phenotypes within each treatment, are presented in the figure. If a p-value is not present it indicates the effect is not significant. P-values for significant effects or trends, across treatments for each immune response phenotype, are presented in the results section.

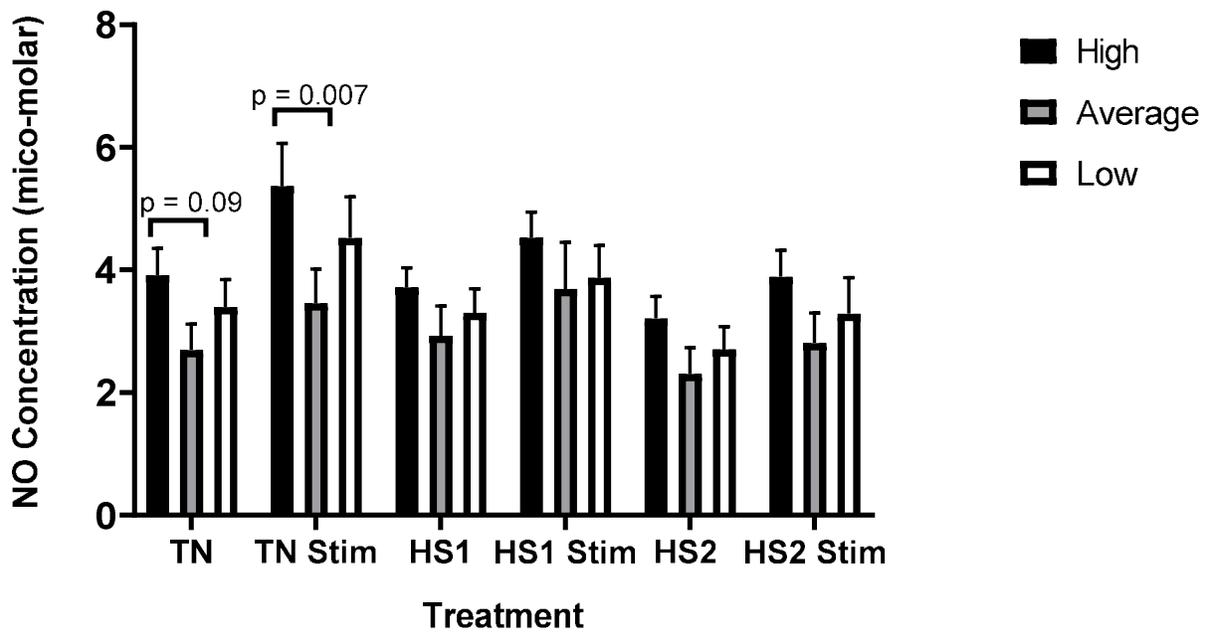

**Figure 2: The production of nitric oxide, by heat treatment for high, average and low immune responders.** Cells were subjected to various heat treatments and nitric oxide production was measured by Griess assay. The data presented are least squared means + standard error of the mean. High immune responders are represented by the black bars, average immune responders by the grey bars and low immune responders by the white bars. P-values for significant effects or trends are presented in the figure. If a p-value is not present it indicates the effect is not significant.

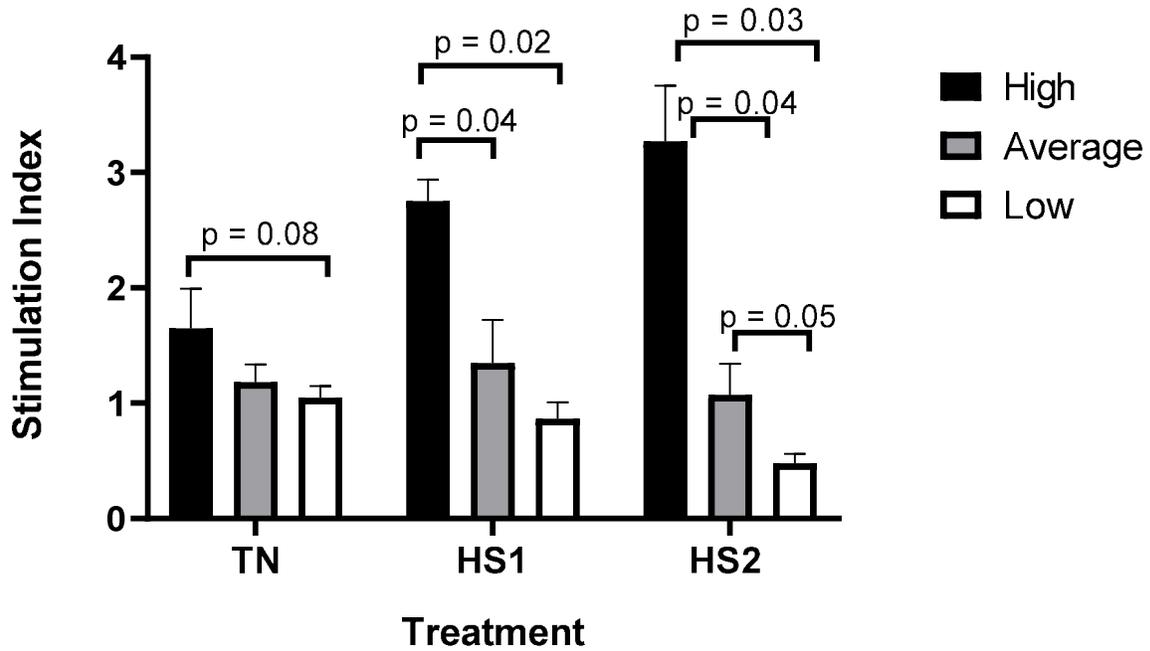

**Figure 3: The cell proliferation, by heat treatment, for high, average and low immune responders.** Cells were subjected to various heat treatments and cell proliferation was measured by MTT assay. The data presented are least squared means + standard error of the mean. High immune responders are represented by the black bars, average immune responders by the grey bars and low immune responders by the white bars. P-values for significant effects or trends are presented in the figure. If a p-value is not present it indicates the effect is not significant.

**Declarations**

*Ethics Approval and Consent to Participate*

This study has been approved by the Animal Care Committee at the University of Guelph under the animal utilization protocol # 3555.

*Consent for Publication*

Not applicable.

*Availability of Data and Materials*

The datasets used and analysed during the current study are available from the corresponding author on reasonable request.

*Competing Interests*

The authors declare that they have no competing interests.

*Funding*

Animal use for this study was supported by an Ontario Ministry of Agriculture and Rural Affairs Tier II grant. Collection and analysis of all the samples was supported by the Canadian First Research Excellence Fund. The Dairy Farmers of Ontario, the Ontario Veterinary College and Food From Thought provided support for the corresponding author to perform interpretation of data and to write the manuscript.

*Author's Information*

Affiliations

**Department of Pathobiology, Ontario Veterinary College, University of Guelph, 50 Stone Road, Guelph, ON, N1G 2W1, Canada**

Shannon L Cartwright, Marnie McKechnie, Julie Schmied, Alexandra M Livernois & Bonnie A Mallard

**Centre for Genetic Improvement of Livestock, Animal Biosciences, University of Guelph, 50 Stone Rd, Guelph, ON, N1G 2W1, Canada**

Alexandra M Livernois & Bonnie A Mallard

**Corresponding Author**

Correspondence to Shannon L Cartwright

*Author's Contributions*

SLC evaluated immune response phenotypes in all cows and calculated EBVs. SLC was also involved in optimization of all assays and collection of blood from all cows as well as performed BMC isolation and analysis of HSP70 concentration, NO production and cell proliferation. SLC completed all statistical analysis and prepared the initial manuscript. MM assisted in blood collection as well as BMC isolation and analysis of HSP70 concentration, NO production and cell proliferation. JS assisted with blood collection as well as BMC isolation and analysis of HSP70 concentration, NO production and cell proliferation. JS also provided both intellectual and technical support in optimization of all assays. AML provided intellectual support in both optimization of assays and statistical analysis. BAM oversaw entire study and provided intellectual support in all aspects of the research. BAM also helped manuscript preparation. All authors' have read and approved the final manuscript.

*Acknowledgements*

The authors would like to thank the staff, especially Laura Wright, at the Elora Research and Innovation Centre for their exceptional care of the cattle involved in this study.

**References**

1. IPCC (Intergovermental Panel on Climate Change). Climate Change 2007: The Physical Science Basis. 2007. doi:10.1256/wea.58.04.

2. Kadzere CT, Murphy MR, Silanikove N, Maltz E. Heat stress in lactating diary cows: a review. J Anim Sci. 2002;77:59–91.

3. Bett B, Kiunga P, Gachohi J, Sindato C, Mbotha D, Robinson T, et al. Effects of climate change on the occurrence and distribution of livestock diseases. Prev Vet Med. 2017;137 November 2015:119–29. doi:10.1016/j.prevetmed.2016.11.019.

4. Salak-Johnson JL, McGlone JJ. Making sense of apparently conflicting data: stress and immunity in swine and cattle. J Anim Sci. 2007;85 13 Suppl:81–8.

5. Das R, Sailo L, Verma N, Bharti P, Saikia J, Imtiwati, et al. Impact of heat stress on health and performance of dairy animals: A review. Veterinary World. 2016;9.

6. Dahl GE, Tao S, Laporta J. Heat Stress Impacts Immune Status in Cows Across the Life Cycle. Front Vet Sci. 2020;7 March:1–15.

7. Bishop-Williams KE, Berke O, Pearl DL, Hand K, Kelton DF. Heat stress related dairy cow mortality during heat waves and control periods in rural Southern Ontario from 2010-2012. BMC Vet Res. 2015.